# The collision frequencies of charged particles in the complex plasmas with the non-Maxwellian velocity distributions

Ma Baojing, Du Jiulin

*Department of Physics, School of Science, Tianjin University, Tianjin 300350, China*

**Abstract** We study the collision frequencies of charged particles in the complex plasmas with the non-Maxwellian velocity distributions. The average collision frequencies of electron-ion, electron-electron and ion-ion are derived in the two-parameter ($r,\sigma$) and the three-parameter ($\alpha, r,\sigma$) velocity distributions, respectively. We show that these average collision frequencies in the complex plasmas depend strongly on the parameters in the non-Maxwellian distributions and thus are significant deviations from those in the plasmas with a Maxwell velocity distribution. Numerical analyses are made of the effects of the parameters on the average collision frequencies. The results have important effect on transport coefficients and their properties of charged particles in the highly ionized complex plasmas with the non-Maxwell distributions.

## 1. Introduction

The collision phenomenon is one of the most basic features of plasma gas dynamics, where the interactions between particles are taking place along with exchanges of momentum or kinetic energy. The nature of plasmas depends on the simultaneous interactions between the particles and the Coulomb forces are often directly involved in the collisions. There are two major collision types depending on whether the Coulomb forces are directly involved in the collision process: one is the non-Coulomb collisions, taking place between neutral particles, and between a neutral particle and a charged particle; the other is the Coulomb collisions taking place between charged particles. The non-Coulomb collisions describe the transient interactions of the particles in the distance equal to the size of the particle in weakly ionized plasmas [1]. The Coulomb collisions can describe the scattering of charged particles in the long-range of Debye length [2]. In both the collisions, there are still the elastic collision and the non-elastic collision [3].

The transport phenomena in plasmas, such as diffusion, heat conduction, and viscosity etc. are closely related to the collision effects [4-9], because the transport coefficients describing the transport processes are generally expressed with the average collision frequencies of the plasmas. The collision frequency is the average number of collisions experienced by each particle unit time, so the average collision frequency usually depends on the velocity distribution function of the particles in the plasmas [10]. Especially, for some nonequilibrium and complex plasmas, because the velocity distribution functions of particles are often not Maxwellian, the average collision frequencies in the plasmas are generally different from those in the Maxwell-distributed plasmas.

Non-Maxwell velocity distributions are universal in nonequilibrium complex systems. In nonextensive statistical mechanics, the complex plasma system being at a nonequilibrium stationary state can be described by a power-law q-distribution [11-14]. In the cases of many astrophysical and space plasmas, the velocity distributions appeared reasonably Maxwellian at low energies but had a 'superthermal' power-law tail at high energies, which were a large deviation from a Maxwellian distribution. The non-Maxwell velocity distributions in the astrophysical and space complex plasmas was observed, and in 1968, Vasyliunas introduced a non-Maxwellian empirical function (the kappa-distribution or Vasyliunas distribution) to fit the magnetic layer plasma measured by OGO1 and OGO3 satellite with high-energy tail characteristics [15]. In the same year, Montgomery et al, based on the observation of solar wind electronics, indicated that the solar wind electron velocity distribution has a high-speed tail of non-Maxwellian velocity distribution. Now nonextensive statistical mechanics has been widely used to research various complex systems, such as astrophysical and space plasmas [16-19], self-gravitating systems [20-23], and chemical reaction systems [24-25]. It has also been employed to study high-energy physics [26], anomalous diffusion [27], and even biological systems [28] etc.

Here we focus on the average collision frequencies in the non-equilibrium complex plasmas with non-Maxwellian velocity distributions, such as those in the astrophysical and space plasmas. Complex plasma with a variety of non-Maxwellian velocity distributions has caused widespread concern in recent years. Especially, the non-Maxwellian velocity distributions have significant effects on transport processes of particles in complex plasmas. Recently, the average collision frequencies of electron-neutral-particle were studied for the weakly ionized plasma with the power-law $q$-distribution in nonexyensive statistics [9], with Vasyliunas–Carins distribution, with the two-parameter ($r,\sigma$) distribution and the three-parameter ($\alpha, r, \sigma$) velocity distribution [29], respectively. For the highly ionized plasmas, the collision frequencies of electron-electron, ion–ion and electron-ion were studied for the power-law $q$-distribution in nonextensive statistics in Ref. [30]. These works have an important impact on our exploration and understanding of the transport properties in complex plasmas.

The highly ionized plasmas can exist in the stellar interiors [31], in the Sun's corona, Sun's atmosphere [32], and in vast regions of interstellar space around the hot stars [33] etc. In this work, we study the average collision frequencies of electron-ion, electron-electron and ion-ion in the complex plasma with the two-parameter ($r, \sigma$) and three-parameter ($\alpha, r, \sigma$) non-Maxwellian distributions, respectively, and then analyze the roles of these parameters in the average collision frequencies.

The paper is organized as follows. In Section 2, we introduce the types of the non-Maxwell velocity distributions. In Section 3, we study the collision frequencies of the charged particles in the complex plasmas with the non-Maxwellian distributions. In Section 4, we make the numerical analyses on the average collision frequencies. Finally in Section 5, we give the conclusion.

**2. The non-Maxwell velocity distributions**

The most basic one of non-Maxwell distributions is the well-known kappa-distribution observed in astrophysical and space plasmas [15], expressed by

$$f_\kappa(\mathbf{v}) = N\left[\frac{2\pi k_B \tilde{T}}{m}\left(\kappa - \frac{3}{2}\right)\right]^{-2/3} \frac{\Gamma(\kappa+1)}{\Gamma\left(\kappa - \frac{1}{2}\right)}\left[1 + \frac{1}{\kappa}\frac{\mathbf{v}^2}{w_\kappa^2}\right]^{-(\kappa+1)}, \qquad (1)$$

where $\mathbf{v}$ is velocity, $k_B$ is Boltzmann constant, $N$ is number density, $m$ is mass of the particle, $\tilde{T}$ is temperature of the plasma, $w_\kappa = w_0[(\kappa-3/2)/\kappa]^{1/2}$ is a characteristic speed with the most probable speed $w_0 = \sqrt{2k_B T/m}$, and $\kappa > 3/2$ is the parameter describing the deviation from a thermal equilibrium. Only when we take $\kappa \to \infty$, the kappa-distribution (1) becomes to a Maxwellian distribution and the system reaches to a thermal equilibrium. Theoretically, the kappa-distribution is found to be exactly equivalent to the power-law $q$-distribution in nonextensive statistics if one makes the parameter transformation [34],

$$2T=(7-5q)\tilde{T} \quad \text{and} \quad (q-1)^{-1}=\kappa+1, \tag{2}$$

where $T$ is the temperature of the $q$-distribution and $q$ is the nonextensive parameter. $\tilde{T}$ is not only the temperature in the kappa-distribution, but it is also the physical temperature of the $q$-distribution in nonextensive statistics. Based on this parameter transformation, the $\kappa$-distributed complex plasmas can be well studied under the framework of nonextensive statistics.

Compared to traditional Maxwellian distribution, the kappa-distribution better fits some of the observation data of the astrophysical and space plasmas, however, more and more complex non-Maxwell velocity distributions had been studied in complex plasmas. In this respect, Summers et al [35] introduced the generalized Lorentzian distribution, Qureshi et al [36] introduced the generalized two-parameter $(r, \sigma)$ distribution and Abid et al introduced the two-parameter $(\alpha, \kappa)$ distribution [37]. Later, these two-parameter distributions were applied to different processing of plasmas, such as the parallel propagating electromagnetic modes of hot magnetized plasma [38], the dust surface potential in the multi-ion dusty plasma [19], and a fit to the data of the magneto sheath electron and the solar wind proton data [39] etc.

The two-parameter $(r,\sigma)$ distribution in nonequilibrium complex plasmas can be written for $j$th plasma species (e.g., electrons and ions) [29, 36] as

$$f_{r,\sigma}(\mathbf{v}_j) = \frac{3E_{r,\sigma}N_j}{4\pi X_{r,\sigma}^{3/2} v_{Tj}^2}\left[1+\frac{1}{\sigma-1}\left(\frac{\mathbf{v}_j^2}{X_{r,\sigma}v_{Tj}^3}\right)^{1+r}\right]^{-\sigma}, \tag{3}$$

where $N_j$ is the particle number density, $\mathbf{v}_j$ is the velocity, $v_{Tj}=(k_B T_j/m_j)^{1/2}$ is the thermal velocity, $T_j$ is temperature, $m_j$ is mass and

$$E_{r,\sigma} = \frac{(\sigma-1)^{\frac{-3}{2r+2}}\Gamma(\sigma)}{\Gamma\left(\sigma-\frac{3}{2r+2}\right)\Gamma\left(1+\frac{3}{2r+2}\right)}, \tag{4}$$

$$X_{r,\sigma} = \frac{3\Gamma\left(\frac{3}{2r+2}\right)\Gamma\left(\sigma-\frac{3}{2r+2}\right)}{(\sigma-1)^{\frac{1}{r+1}}\Gamma\left(\sigma-\frac{5}{2r+2}\right)\Gamma\left(\frac{5}{2r+2}\right)}. \tag{5}$$

The two parameters $r$ and $\sigma$ have significantly determined deviation of the plasma from the thermal equilibrium. When we take $r\to 0$ and $\sigma=\kappa+1$, the $(r,\sigma)$ distribution (3) reduces to the kappa-distribution function, when we take $r\to 0$, $\sigma\to\infty$, the $(r,\sigma)$ distribution (3) becomes a Maxwellian distribution. Based on the normalization and the definition of temperature in the above distribution function, the parameters $r$ and $\sigma$ must satisfy the conditions: $\sigma>1$ and $\sigma(1+r)>5/2$. Generally, if we increase the value of $r$ only and keep the value of $\sigma$ unchanged, the contribution of high energy particles decreases, but the shoulder of the distribution curve becomes wider. Similarly, if we increase the value of $\sigma$ only and keep the value of $r$ fixed, the result is the same [37].

A large number of studies have shown that there are many linear and nonlinear space plasma phenomena that can be explained by non-Maxwellian distributions of plasmas, such as those in the solar wind [40], in the magnetocaloric [39] and auroral regions [41,42]. And at the same time, people are looking for more generalized distribution functions to understand the space plasma physics. In 2017, a three-parameter $(\alpha, r, \sigma)$ velocity distribution was introduced [43]: $\alpha$ is the rate of energetic particles on the shoulder, $r$ is the energetic particles on a broad shoulder, $\sigma$ is the superthermality on the tail of velocity distribution curve of plasma species. The three-parameter distribution can understand the observed space plasma phenomena more accurately.

The three-parameter ($\alpha$, $r$, $\sigma$) velocity distribution function is a more complex non-Maxwell distribution, which can be expressed for the $j$th plasma species [29,43,44] as

$$f_{\alpha,r,\sigma}(\mathbf{v}_j) = \frac{3N_j}{4\pi v_{Tj}^3} \frac{\rho_{\alpha,r,\sigma}}{X_{r,\sigma}^{3/2}} \left(1 + \alpha \frac{\mathbf{v}_j^4}{v_{Tj}^4}\right) \left[1 + \frac{1}{\sigma-1}\left(\frac{\mathbf{v}_j^2}{X_{r,\sigma}v_{Tj}^2}\right)^{1+r}\right]^{-\sigma}, \quad (6)$$

where [44], 
$$\rho_{\alpha,r,\sigma} = \frac{\Gamma(\sigma)}{(1+9\eta_{r,\sigma}\alpha)(\sigma-1)^{\frac{3}{2r+2}}\Gamma\left(\sigma-\frac{3}{2r+2}\right)\Gamma\left(1+\frac{3}{2r+2}\right)}, \quad (7)$$

$$\eta_{r,\sigma} = \frac{\Gamma\left(\sigma-\frac{3}{2r+2}\right)\Gamma\left(\frac{3}{2r+2}\right)\Gamma\left(\sigma-\frac{7}{2r+2}\right)\Gamma\left(\frac{7}{2r+2}\right)}{\Gamma^2\left(\sigma-\frac{5}{2r+2}\right)\Gamma^2\left(\frac{5}{2r+2}\right)}. \quad (8)$$

This velocity distribution function has three parameters $\alpha$, $r$ and $\sigma$, satisfying the condition $\sigma > 1$, $\alpha > 0$ and $\sigma(1+r) > 5/2$. Actually, the three-parameter distribution function (6) consists of many different distribution functions. For example, when we take $\alpha = 0$, it becomes the two-parameter ($r,\sigma$) distribution, when we take $\alpha \to 0$ and $r \to 0$, it reduces to the $\sigma$-distribution in nonextensive statistics, when we take $r \to 0$ and $\sigma \to \infty$, it reduces to Cairns-distribution [45], when we take $r \to 0$ and $\sigma = \kappa+1$, it reduces to Vasyliunas-Cairns distribution [46], when we take $\alpha \to 0$, $r \to 0$ and $\sigma = \kappa+1$, it becomes the kappa-distribution, and when we take $\alpha \to 0$, $r \to 0$ and $\sigma \to \infty$, it recovers to a Maxwellian distribution.

## 3. The collision frequencies of charged particles in the complex plasmas

The effective radius of the electrostatic interactions at normal temperatures is about hundred times that of the collisions between a charged particle and a neutral particle, so the Coulomb collisions play a main role in the highly ionized plasma [47]. When the incident charged particle is deflected by the Coulomb collisions of target charged particles, small-angle scatterings produce a superposed 90-degree net deflection of the incident particle long before the deflection of a single large-angle close collision, therefore the collision frequency of charged particles in the highly ionized plasma needs to be calculated on the basis of the Coulomb scattering model. Since Coulomb electrostatic interactions between charged particles are long-range interactions, the Coulomb scattering model is effectively applied to highly ionized plasma studies.

The Coulomb collision in the highly ionized plasma consists of two different transport properties: the velocity space transport and the physical space transport. In the velocity space transport, the momentum exchange and the energy exchange correspond to the different Coulomb cross sections, respectively [2]. When the charged particles are incident in the plasma, the momentum relaxation of the incident particles represents the total loss of the incident particle directional momentum. The test particle method can be used to calculate the momentum relaxation collision frequency of the incident tape electric particles, wherein the Coulomb cross section of the tested particles is a function of the relative velocity of the test particles and the target particles [1]. In fact, the test particles in the highly ionized plasma are not affected by the target particle electric field of the distance exceeding Debye length [2]. For the momentum relaxation of the test particle $j$, the momentum relaxation collision frequency in the background of field charged particles $k$ can be expressed [3,48] as

$$\nu_{jk} = \frac{N_k Z_j^2 Z_k^2 e^4 \ln \Lambda}{4\pi \varepsilon_0^2 m_{jk}^2 v_{jk}^3}, \tag{9}$$

where $m_{jk} = m_j m_k / (m_j + m_k)$ is the reduced mass, $v_{jk} = |\mathbf{v}_j - \mathbf{v}_k|$ is an absolute value of the relative velocity, $Z_j e$ is the charge of an $j$-particle, $Z_k e$ is the charge of an $k$-particle, $\varepsilon_0$ is the vacuum dielectric constant, $N_k$ is the number density of $k$-particles, and $\ln\Lambda$ is the Coulomb scattering factor determined by the Debye length. In Eq. (9), the collision frequency $\nu_{jk}$ is a function of the relative velocity $v_{jk}$, and therefore, the average collision frequency will depend on the velocity distribution of charged particles in the plasma under consideration. Namely, the average collision frequency is that

$$\overline{\nu}_{jk} = \frac{N_k Z_j^2 Z_k^2 e^4 \ln \Lambda}{4\pi \varepsilon_0^2 m_{jk}^2} \langle v_{jk}^{-3} \rangle, \tag{10}$$

In the conventional statistical physics framework, the charged particles are in the thermodynamic equilibrium state and have a Maxwellian velocity distribution, so the average collision frequency of electron-ion collision is calculated [3] as

$$\overline{\nu}_{ei} = \frac{\sqrt{2} N_i Z_i^2 e^4 \ln \Lambda}{12 \varepsilon_0^2 m_e^{1/2} (\pi k_B T_e)^{3/2}}, \tag{11}$$

and the average collision frequency of electron-electron collision or ion-ion collision is calculated as

$$\overline{\nu}_{jj} = \frac{N_j Z_j^4 e^4 \ln \Lambda}{12 \varepsilon_0^2 m_j^{1/2} (\pi k_B T_e)^{3/2}}, \quad j = e, i. \tag{12}$$

Eq. (11) and Eq. (12) show that the average collision frequency of the electron-ion collision is different from the average collision frequency of electron-electron, ion-ion collision in a Maxwellian velocity distribution. But for nonequilibrium complex plasmas, the system is not in the thermal equilibrium state and the velocity distributions are not Maxwellian one, the average collision frequency needs to be recalculated on the basis of the velocity distribution functions of charged particles.

**3.1.1 The average collision frequency of electron-ion in the two-parameter ($r$, $\sigma$) distribution**

According to Eq. (10), the average collision frequency of electron–ion in the plasma with the two-parameter ($r$, $\sigma$) distribution is calculated by

$$\overline{\nu}_{ei}(r,\sigma) = \frac{N_i Z_i^4 e^4 \ln \Lambda}{4\pi \varepsilon_0^2 m_{ei}^2} \langle v_{ei}^{-3} \rangle_{r,\sigma}. \tag{13}$$

In Eq. (13), the integration calculation on the average value $\langle v_{ei}^{-3} \rangle_{r,\sigma}$ of the relative velocity of electron-ion collision in the two-parameter ($r$, $\sigma$) distribution is divergent. In order to calculate the average collision frequency of electron-ion collision, we use the interaction force exerted by an ion on an electron in the center-of-mass frame when the (light) electron impact (heavy) ion [3],

$$F_{ei} = -N_e m_{ei} \langle \nu_{ei} \mathbf{v}_{ei} \rangle_{r,\sigma}, \tag{14}$$

where $m_{ei} = m_e m_i / (m_e + m_i) \approx m_e$ is the reduced mass, and $\mathbf{v}_{ei} = \mathbf{v}_e - \mathbf{v}_i \approx \mathbf{v}_e$ is the relative velocity vector, and the average value is defined for the charged particles following the two-parameter ($r$, $\sigma$) distribution [49] as

$$\langle \nu_{ei} \mathbf{v}_{ei} \rangle_{r,\sigma} = \int \nu_{ei} \mathbf{v}_{ei} f_{r,\sigma} d\mathbf{v} / \int f_{r,\sigma} d\mathbf{v}. \tag{15}$$

Generally, if we assume that the drifting electrons have a non-zero mean velocity $\mathbf{u}$, the two-parameter $(r,\sigma)$ distribution function (3) of the electrons can be written as

$$f_{r,\sigma}(\mathbf{v}_e) = \frac{3E_{r,\sigma}N_e}{4\pi X_{r,\sigma}^{3/2} v_{Te}^3}\left[1 + \frac{1}{\sigma-1}\left(\frac{(\mathbf{v}_e - \mathbf{u})^2}{X_{r,\sigma} v_{Te}^2}\right)^{1+r}\right]^{-\sigma}, \tag{16}$$

Since the non-zero mean velocity vector **u** has no impact on the average collision frequency [3], to simplify the calculations, we take $\mathbf{u} = (0,0,u_z)$, where $u_z$ is very small compared with the thermal speed, namely $u_z \ll (k_B T_e / m_e)^{1/2}$. Then Eq. (16) can be expended for **u**, up to the first order term of **u** as

$$f_{r,\sigma}(\mathbf{v}_e) \approx \frac{3E_{r,\sigma}N_e}{4\pi (X_{r,\sigma})^{3/2} v_{Te}^3}\left\{1 + \frac{2\sigma(1+r)\mathbf{v}_e \cdot u}{(\sigma-1)X_{r,\sigma}v_{Te}^2}\left(\frac{\mathbf{v}_e^2}{X_{r,\sigma}v_{Te}^2}\right)^r \left[1 + \frac{1}{\sigma-1}\left(\frac{\mathbf{v}_e^2}{X_{r,\sigma}v_{Te}^2}\right)^{1+r}\right]^{-1}\right\}$$
$$\left[1 + \frac{1}{\sigma-1}\left(\frac{\mathbf{v}_e^2}{X_{r,\sigma}v_{Te}^2}\right)^{1+r}\right]^{-\sigma}. \tag{17}$$

According to Eq. (9), the collision frequency becomes

$$\nu_{ei} = \frac{N_i Z_i^2 e^4 \ln\Lambda}{4\pi\varepsilon_0^2 m_{ei}^2 v_{ei}^3} \approx \frac{N_i Z_i^2 e^4 \ln\Lambda}{4\pi\varepsilon_0^2 m_e^2 v_e^3}, \tag{18}$$

Substituting Eqs. (17) and (18) into Eq. (15), we can derive (see Appendix A) that

$$F_{ei} = -\frac{N_e N_i Z_i^2 e^4 u_z m_e^{1/2} \ln\Lambda}{4\pi\varepsilon_0^2 (k_B T_e)^{3/2}} \frac{E_{r,\sigma}}{X_{r,\sigma}^{3/2}}, \tag{19}$$

On the other hand, Eq. (14) is equal to

$$F_{ei} = -N_e m_e \langle \nu_{ei} \rangle_{r,\sigma} u_z, \tag{20}$$

Comparing Eq. (19) with Eq. (20), we find the average collision frequency of electron-ion collisions,

$$\bar{\nu}_{ei}(r,\sigma) = \langle \nu_{ei} \rangle_{r,\sigma} = \frac{N_i Z_i^2 e^4 \ln\Lambda}{4\pi\varepsilon_0^2 m_e^{1/2}(k_B T_e)^{3/2}} \frac{E_{r,\sigma}}{X_{r,\sigma}^{3/2}}. \tag{21}$$

It is clear that the average collision frequency of electron-ion collisions is strongly dependent on the two-parameter $r$ and $\sigma$, and so is significantly different from the result Eq. (11) in the plasma with a Maxwellian velocity distribution. Only when we take the limit $r \to 0$ and $\sigma \to \infty$, Eq. (21) becomes Eq. (11).

### 3.1.2 The average collision frequencies of electron-electron and ion-ion in the two-parameter (*r*, σ) distribution

According to Eq. (10), the average collision frequencies of electron–electron and ion–ion collisions in the plasma with the two-parameter $(r,\sigma)$ distribution are calculated by

$$\bar{\nu}_{jj}(r,\sigma) = \langle \nu_{jj} \rangle_{r,\sigma} = \frac{N_j Z_j^4 e^4 \ln\Lambda}{4\pi\varepsilon_0^2 m_{jj}^2}\langle v_{jj}^{-3}\rangle_{r,\sigma}, j=e,i. \tag{22}$$

For the same reason as that in Eq. (13), the average collision frequencies can be derived by means of calculating the interaction force between the same kinds of charged particles. The interaction force in the center-of-mass frame can be given [3] as

$$F_{jj} = -N_j m_{jj}\langle \nu_{jj}\mathbf{v}_{jj}\rangle_{r,\sigma}, \tag{23}$$

According to Eq. (9), the collision frequency is written as

$$\nu_{jj} = \frac{N_j Z_j^4 e^4 \ln \Lambda}{4\pi\varepsilon_0^2 m_{jj}^2 v_{jj}^3} \approx \frac{N_j Z_j^4 e^4 \ln \Lambda}{\pi\varepsilon_0^2 m_j^2 |v_{j,1} - v_{j,2}|^3}, \tag{24}$$

According to Eq. (3), the two-parameter $(r,\sigma)$ distribution function of the pair of charged particles is taken as

$$f_{r,\sigma}(\mathbf{v}_j) = \frac{3 E_{r,\sigma} N_j}{4\pi X_{r,\sigma}^{3/2} v_{Tj}^3} \left[ 1 + \frac{1}{\sigma-1} \left( \frac{(\mathbf{v}_j - \mathbf{u})^2}{X_{r,\sigma} v_{Tj}^2} \right)^{1+r} \right]^{-\sigma}, \tag{25}$$

In the same way to calculate the interaction force in Eq. (23) as that in of Eq. (14), we can derive that

$$F_{jj} = -\frac{N_j^2 Z_j^4 e^4 u_z m_e^{1/2} \ln \Lambda}{4\sqrt{2}\pi\varepsilon_0^2 (k_B T_e)^{3/2}} \frac{E_{r,\sigma}}{X_{r,\sigma}^{3/2}}, \tag{26}$$

On the other hand, because the mean velocity vector $\mathbf{u}$ is along z-axis, the interaction force [3] becomes

$$F_{jj} = -N_j m_j \langle \nu_{jj} \rangle_{r,\sigma} u_z, \tag{27}$$

Comparing Eq. (26) with Eq. (27), we obtain the average collision frequencies of electron-electron and ion-ion collisions ($j = e, i$),

$$\bar{\nu}_{jj}(r,\sigma) = \langle \nu_{jj} \rangle_{r,\sigma} = \frac{N_j Z_j^4 e^4 \ln \Lambda}{4\sqrt{2}\pi\varepsilon_0^2 m_j^{1/2} (k_B T_e)^{3/2}} \frac{E_{r,\sigma}}{X_{r,\sigma}^{3/2}}. \tag{28}$$

Thus, the average collision frequencies given in Eq. (28) are strongly dependent on the two-parameter $r$ and $\sigma$. When we take the limit $r \to 0$ and $\sigma \to \infty$, (28) becomes into $\bar{\nu}_{jj}$ in Eq. (12).

## 3.2.1 The average collision frequency of electron-ion in the three-parameter ($\alpha$, $r$, $\sigma$) distribution

According to Eq. (10), the average collision frequency of electron–ion in the plasma with the three-parameter ($\alpha, r, \sigma$) distribution is calculated by

$$\bar{\nu}_{ei}(\alpha, r, \sigma) = \frac{N_i Z_i^4 e^4 \ln \Lambda}{4\pi\varepsilon_0^2 m_{ei}^2} \langle v_{ei}^{-3} \rangle_{\alpha,r,\sigma}. \tag{29}$$

The interaction force exerted by an ion on an electron in the center-of-mass frame can be given as

$$F_{ei} = -N_e m_{ei} \langle \nu_{ei} \mathbf{v}_{ei} \rangle_{\alpha,r,\sigma}, \tag{30}$$

where the average value defined [49] as

$$\langle \nu_{ei} \mathbf{v}_{ei} \rangle_{\alpha,r,\sigma} = \int \nu_{ei} \mathbf{v}_{ei} f_{\alpha,r,\sigma} d\mathbf{v} / \int f_{\alpha,r,\sigma} d\mathbf{v}. \tag{31}$$

Generally, we still assume the drifting electrons have a non-zero but small mean velocity $\mathbf{u} = (0,0,u_z)$, accordingly, the three-parameter $(\alpha, r, \sigma)$ velocity distribution function (5) of the electrons can be written as

$$f_{\alpha,r,\sigma}(\mathbf{v}_e) = \frac{3 N_e}{4\pi v_{Te}^3} \frac{\rho_{\alpha,r,\sigma}}{X_{r,\sigma}^{3/2}} \left( 1 + \alpha \frac{(v_e - \mathbf{u})^4}{v_{Te}^4} \right) \left[ 1 + \frac{1}{\sigma-1} \left( \frac{(\mathbf{v}_e - \mathbf{u})^2}{X_{r,\sigma} v_{Te}^2} \right)^{1+r} \right]^{-\sigma}, \tag{32}$$

For convenience, we use the abbreviations,

$$A = 1 + \alpha \frac{v_e^4}{v_{Te}^4}, \quad B = 1 + \frac{1}{\sigma-1} \left( \frac{v_e^2}{X_{r,\sigma} v_{Te}^2} \right)^{1+r}.$$

And then Eq. (32) can be expended up to the first order terms of $\mathbf{u}$ as

$$f_{\alpha,r,\sigma}(\mathbf{v}_e) = \frac{3N_e \rho_{\alpha,r,\sigma}}{4\pi X_{r,\sigma}^{3/2} v_{Te}^3} AB^{-\sigma} \left[ 1 + \mathbf{u}B^{-1} \left( \frac{\mathbf{v}_e^2}{X_{r,\sigma} v_{Te}^2} \right)^r \frac{2\sigma(1+r)}{X_{r,\sigma}(\sigma-1)} - \mathbf{u}A^{-1} \frac{4\alpha \mathbf{v}_e^2}{v_{Te}^4} \right], \quad (33)$$

Substituting Eqs. (18) and (33) into Eq. (31), we can derive (see Appendix B) that

$$F_{ei} = -\frac{N_e N_i Z_i^2 e^4 u_z m_e^{1/2} \ln \Lambda}{4\pi \varepsilon_0^2 (k_B T_e)^{3/2}} \frac{\rho_{\alpha,r,\sigma}}{X_{r,\sigma}^{3/2}}, \quad (34)$$

On the other hand, Eq. (30) is equal to that

$$F_{ei} = -N_e m_e \langle v_{ei} \rangle_{\alpha,r,\sigma} u_z \quad (35)$$

Comparing Eq. (34) with Eq. (35), we find the average collision frequency of electron-ion collisions,

$$\overline{v}_{ei}(\alpha, r, \sigma) = \langle v_{ei} \rangle_{\alpha,r,\sigma} = \frac{N_i Z_i^2 e^4 \ln \Lambda}{4\pi \varepsilon_0^2 m_i^{1/2} (k_B T_e)^{3/2}} \frac{\rho_{\alpha,r,\sigma}}{X_{r,\sigma}^{3/2}}. \quad (36)$$

Thus, the average collision frequency given in Eq. (36) is strongly dependent on the three-parameter $(\alpha, r, \sigma)$. It is easily proved that when we take $\alpha \to 0$, Eq. (36) becomes $\overline{v}_{ei}(r,\sigma)$ in Eq.(21), when we take $\alpha \to 0$, $r \to 0$ and $\sigma \to 1$, Eq. (36) becomes $\overline{v}_{ei}$ in Eq. (11).

**3.2.2 The average collision frequencies of electron-electron and ion-ion in the three-parameter ($\alpha, r, \sigma$) distribution**

According to Eq. (10), the average collision frequencies of electron–electron and ion–ion collisions in the plasma with the three-parameter $(\alpha, r, \sigma)$ distribution are calculated by

$$\overline{v}_{jj}(\alpha, r, \sigma) = \langle v_{jj} \rangle_{\alpha,r,\sigma} = \frac{N_j Z_j^4 e^4 \ln \Lambda}{4\pi \varepsilon_0^2 m_{jj}^2} \langle v_{jj}^{-3} \rangle_{\alpha,r,\sigma}, j = e, i \quad (37)$$

For the same reason as that in Eq. (13), the average collision frequencies are derived by means of calculating the interaction force between the same kinds of charged particles. The interaction force in the center-of-mass frame can be given [3] as

$$F_{jj} = -N_j m_{jj} \langle v_{jj} \mathbf{v}_{jj} \rangle_{\alpha,r,\sigma}. \quad (38)$$

On the basis of Eq. (9), the collision frequencies are written as

$$v_{jj} = \frac{N_j Z_j^4 e^4 \ln \Lambda}{4\pi \varepsilon_0^2 m_{jj}^2 v_{jj}^3} \approx \frac{N_j Z_j^4 e^4 \ln \Lambda}{\pi \varepsilon_0^2 m_j^2 |v_{j,1} - v_{j,2}|^3}. \quad (39)$$

According to Eq. (1), the three-parameter $(\alpha, r, \sigma)$ velocity distribution function of the pair of charged particles is taken as

$$f_{\alpha,r,\sigma}(\mathbf{v}_j) = \frac{3N_j}{4\pi v_{Tj}^3} \frac{\rho_{\alpha,r,\sigma}}{X_{r,\sigma}^{3/2}} \left( 1 + \alpha \frac{(\mathbf{v}_j - \mathbf{u})^4}{v_{Tj}^4} \right) \left[ 1 + \frac{1}{\sigma-1} \left( \frac{(\mathbf{v}_j - \mathbf{u})^2}{X_{r,\sigma} v_{Tj}^2} \right)^{1+r} \right]^{-\sigma}. \quad (40)$$

In the same way to calculate the interaction force using Eq. (40) as that in Eq. (14), we can derive that

$$F_{jj} = -\frac{N_j^2 Z_j^4 e^4 u_z m_j^{1/2} \ln \Lambda}{4\sqrt{2}\pi \varepsilon_0^2 (kT_e)^{3/2}} \frac{\rho_{\alpha,r,\sigma}}{X_{r,\sigma}^{3/2}}. \quad (41)$$

On the other hand, Eq. (41) is equal to that

$$F_{jj} = -N_j m_j \langle v_{jj} \rangle_{\alpha,r,\sigma} u_z, \tag{42}$$

Comparing Eq. (41) with Eq. (42), we find the average collision frequencies of the electron-electron and ion-ion collisions are

$$\overline{v}_{jj}(\alpha,r,\sigma) = \langle v_{jj} \rangle_{\alpha,r,\sigma} = \frac{N_j Z_j^2 e^4 \ln \Lambda}{4\sqrt{2}\pi\varepsilon_0^2 m_j^{1/2} (kT_e)^{3/2}} \frac{\rho_{\alpha,r,\sigma}}{X_{r,\sigma}^{3/2}}. \tag{43}$$

From equation (43) we can obtain the conclusions that when we take $\alpha \to 0$, Eq. (43) becomes $\overline{v}_{jj}(r,\sigma)$ in Eq. (28), and when we take limit $\alpha \to 0$, $r \to 0$ and $\sigma \to 1$, (43) becomes $\overline{v}_{jj}$ in Eq. (12).

## 4. Numerical analysis

In order to show the effects more clearly of these parameters on the average collision frequencies of charged particles in the complex plasmas with the non-Maxwellian velocity distributions and, at the same time, compare with the cases with a Maxwellian velocity distribution, now we make numerical analyses on the average collision frequencies of the electron-ion, the ion-ion and the electron-electron, respectively. For this purpose, by using Eq. (21), Eq. (28), Eq. (11) and Eq. (12), we can write the average collision frequencies in the two-parameter $(r,\sigma)$ distribution as the following equations,

$$\frac{\overline{v}_{ei}(r,\sigma)}{\overline{v}_{ei}} = \frac{\overline{v}_{jj}(r,\sigma)}{\overline{v}_{jj}} = 3\sqrt{\frac{\pi}{2}} \frac{E_{r,\sigma}}{X_{r,\sigma}^{3/2}}, \quad j = e,i \tag{44}$$

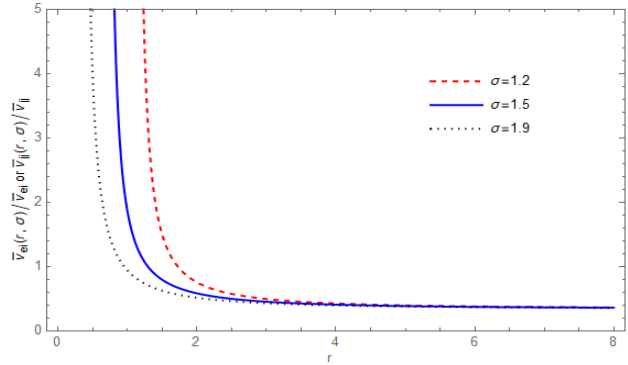

(a)

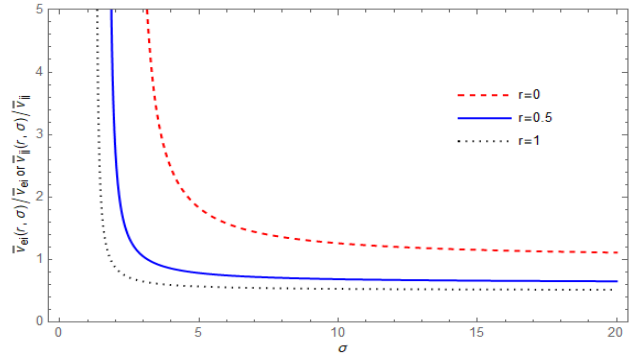

(b)

Figure 1. Dependences of the average collision frequencies (44) on the parameters $r$ and $\sigma$

Based on Eq. (44), we have made the numerical analyses. In Fig. 1, we have shown the role of the parameters $r$ and $\sigma$ in the average collision frequencies of electron-ion, ion-ion and electron- electron in the highly ionized plasma with the two-parameter $(r,\sigma)$ distribution. In Fig. 1(a), we have shown Eq. (44) as a function of the parameter $r$ for three different values of the parameter $\sigma$ (=1.2, 1.5 and 1.9). It is shown that with increase of the parameter $r$, the frequencies $\bar{v}_{ei}(r,\sigma)$ decrease rapidly when $r$ is small and then gradually tend to the same constant for the three different $\sigma$. In Fig. 1(b), we have shown Eq. (44) as a function of the parameter $\sigma$ for three different values of the parameter $r$ (=0, 0.5 and 1). It is shown that with increase of the parameter $\sigma$, the frequencies $\bar{v}_{ei}(r,\sigma)$ decreases rapidly when $\sigma$ is small and then gradually tend to constants for the three different parameters $r$, respectively.

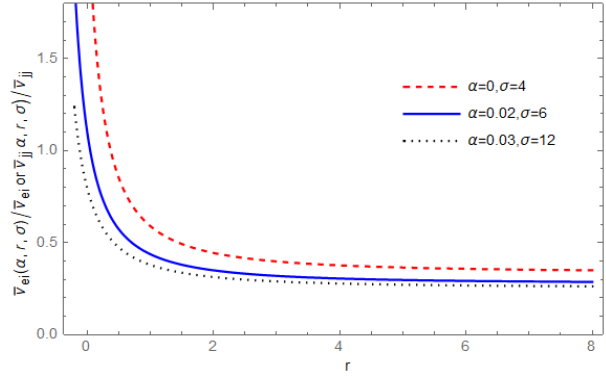

(a)

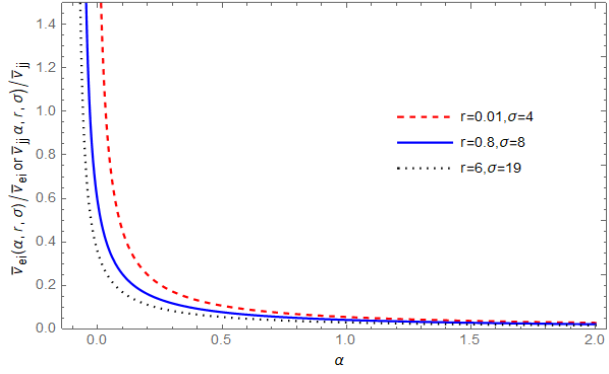

(b)

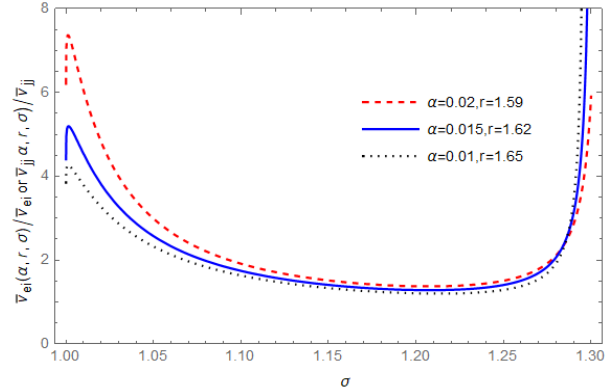

(c)

Figure 2. Dependences of the average collision frequencies (45) on the parameters $\alpha$, $r$ and $\sigma$.

For the average collision frequencies in the three parameter ($\alpha$, $r$, $\sigma$) distribution, by using Eq. (36), Eq. (43), Eq. (11) and Eq. (12), we can write the average collision frequencies as the following equations,

$$\frac{\bar{v}_{ei}(\alpha,r,\sigma)}{\bar{v}_{ei}} = \frac{\bar{v}_{jj}(\alpha,r,\sigma)}{\bar{v}_{jj}} = 3\sqrt{\frac{\pi}{2}} \frac{\rho_{\alpha,r,\sigma}}{X_{r,\sigma}^{3/2}}. \tag{45}$$

Based on Eq. (45), in Fig.2 we have shown dependence of the average collision frequencies on the three parameters $\alpha$, $r$ and $\sigma$. In Fig. 2(a), we give Eq. (45) as a function of the parameter $r$ for three different parameters $\alpha$ (=0, 0.02, 0.03) and $\sigma$ (=4, 6, 12), respectively. In Fig. 2(b), we give Eq. (45) as a function of the parameter $\alpha$ for three different parameters $r$ (=0.01, 0.8, 6) and $\sigma$ (=4, 8, 19), respectively. From Figs. 2(a) and 2(b), we show that with increase of the parameters $r$ and $\alpha$, the frequencies $\bar{v}_{ei}(\alpha,r,\sigma)$ decrease rapidly when $r$ and $\alpha$ are small and then gradually tend to a constant for the other two sets of the parameters with three different values, respectively.

In Fig.2(c), we give Eq. (45) as a function of the parameter $\sigma$ for three different parameters $\alpha$ (=0.02, 0.015, 0.01) and $r$ (=1.59, 1.62, 1.65), respectively, when the parameter $\sigma$ is taken in a small value range 1~1.3. We show that in this value range of $\sigma$, changes of the frequencies $\bar{v}_{ei}(\alpha,r,\sigma)$ are complex. With increase of the parameter $\sigma$, $\bar{v}_{ei}(\alpha,r,\sigma)$ increases, reach a peak and then decrease gradually when $\sigma$ is near 1, but when $\sigma$ is near to 1.3, it increases rapidly, for the other two sets of the parameters $\alpha$ and $r$ with three different values, respectively.

## 5. Conclusion

In summary, we have studied the average collision frequencies of the charged particles in the complex plasmas with the two types of non-Maxwell velocity distributions, namely, the two-parameter ($r$, $\sigma$) distribution and the three-parameter ($\alpha$, $r$, $\sigma$) distribution. The expressions of the average collision frequencies of electron-ion, electron-electron and ion-ion in the plasma with the two types of non-Maxwell velocity distributions are derived respectively. The average collision frequencies for the charged particles in the two-parameter ($r$, $\sigma$) distribution are given by Eqs. (21) and (28) respectively, and the average collision frequencies in the three-parameter ($\alpha$, $r$, $\sigma$) distribution are given by Eqs. (36) and (43), respectively. Numerical analyses have been made to show the effects of these parameters on the average collision frequencies.

We find that all the average collision frequencies in the complex plasmas depend strongly on the two-parameters ($r$, $\sigma$) and the three-parameters ($\alpha$, $r$, $\sigma$) in the non-Maxwell velocity distributions and thus show their significant deviations from those in the plasma with a Maxwellian velocity distribution. These results will have important effects on the transport coefficients and their properties of charged particles in the highly ionized complex plasmas with the non-Maxwell distributions.

**Acknowledgements**

This work is supported by the National Natural science foundation of China under Grant No. 11775156.

**Appendix A**

Eq. (19) is calculated as follows.

$$F_{ei} = -\frac{3N_e N_i Z_i^2 e^4 \ln\Lambda}{16\pi^2\varepsilon_0^2 m_e} \frac{E_{r,\sigma}}{X_{r,\sigma}^{3/2} v_{Te}^3}$$

$$\int d\mathbf{v}\frac{\mathbf{v}}{v^3}\left\{1+\frac{2\sigma(1+r)\mathbf{v}\cdot\mathbf{u}}{(\sigma-1)X_{r,\sigma}v_{Te}^2}\left(\frac{\mathbf{v}^2}{X_{r,\sigma}v_{Te}^2}\right)^r\left[1+\frac{1}{\sigma-1}\left(\frac{\mathbf{v}^2}{X_{r,\sigma}v_{Te}^2}\right)^{1+r}\right]^{-1}\right\}\left[1+\frac{1}{\sigma-1}\left(\frac{\mathbf{v}^2}{X_{r,\sigma}v_{Te}^2}\right)^{1+r}\right]^{-\sigma}$$

$$= -\frac{3N_e N_i Z_i^2 e^4 \ln\Lambda}{8\pi^2\varepsilon_0^2 k_B T_e}\frac{E_{r,\sigma}}{X_{r,\sigma}^{5/2} v_{Te}^3}\frac{\sigma(1+r)}{\sigma-1}\int d\mathbf{v}\frac{\mathbf{v}\mathbf{v}\cdot\mathbf{u}}{v^3}\left(\frac{\mathbf{v}^2}{X_{r,\sigma}v_{Te}^2}\right)^r\left[1+\frac{1}{\sigma-1}\left(\frac{\mathbf{v}^2}{X_{r,\sigma}v_{Te}^2}\right)^{1+r}\right]^{-\sigma-1}. \quad (A.1)$$

Because of $\mathbf{u}=(0,0,u_z)$, we obtain that

$$F_{ei} = -\frac{3N_e N_i Z_i^2 e^4 u_z \ln\Lambda}{8\pi^2\varepsilon_0^2 k_B T_e}\frac{E_{r,\sigma}}{X_{r,\sigma}^{5/2} v_{Te}^3}\frac{\sigma(1+r)}{\sigma-1}$$

$$\int_{-v_{\max}}^{v_{\max}} d\mathbf{v}\frac{v_z^2}{v^3}\left(\frac{\mathbf{v}^2}{X_{r,\sigma}v_{Te}^2}\right)^r\left[1+\frac{1}{\sigma-1}\left(\frac{\mathbf{v}^2}{X_{r,\sigma}v_{Te}^2}\right)^{1+r}\right]^{-\sigma-1}$$

$$= -\frac{N_e N_i Z_i^2 e^4 u_z \ln\Lambda}{2\pi\varepsilon_0^2 k_B T_e}\frac{\sigma(1+r)E_{r,q}}{(\sigma-1)X_{r,q}^{5/2}v_{Te}^3}\int_0^{v_{\max}} dv\left(\frac{v^2}{X_{r,\sigma}v_{Te}^2}\right)^r v\left[1+\frac{1}{\sigma-1}\left(\frac{v^2}{X_{r,\sigma}v_{Te}^2}\right)^{1+r}\right]^{-\sigma-1}. \quad (A.2)$$

When we have $v_{\max}=\infty$, we obtain that

$$F_{ei} = -\frac{N_e N_i Z_i^2 e^4 u_z m_e^{1/2} \ln\Lambda}{4\pi\varepsilon_0^2(k_B T_e)^{3/2}}\frac{E_{r,\sigma}}{X_{r,\sigma}^{3/2}}. \quad (A.3)$$

**Appendix B**

Eq. (34) is calculated as follows.

$$F_{ei} = -\frac{3N_e N_i Z_i^2 e^4 \ln\Lambda}{16\pi^2\varepsilon_0^2 m_e}\frac{\rho_{\alpha,r,\sigma}}{X_{r,\sigma}^{3/2}v_{Te}^3}\int d\mathbf{v}\frac{\mathbf{v}}{v^3}\left\{1+\mathbf{u}B^{-1}\left(\frac{\mathbf{v}_e^2}{X_{r,\sigma}v_{Te}^2}\right)^r\frac{2\sigma(1+r)}{X_{r,\sigma}(\sigma-1)}-\mathbf{u}A^{-1}\frac{4\alpha\mathbf{v}_e^2}{v_{Te}^4}\right\}AB^{-\sigma}$$

$$= -\frac{3N_e N_i Z_i^2 e^4 \ln\Lambda}{16\pi^2\varepsilon_0^2 k_B T_e}\frac{\rho_{\alpha,r,\sigma}}{X_{r,\sigma}^{3/2}v_{Te}^3}\int d\mathbf{v}\frac{\mathbf{v}\mathbf{v}\cdot\mathbf{u}}{v^3}\left[AB^{-\sigma-1}\left(\frac{\mathbf{v}^2}{X_{r,\sigma}v_{Te}^2}\right)^r\frac{2\sigma(1+r)}{X_{r,\sigma}(\sigma-1)}-\frac{4\alpha\mathbf{v}_e^2}{v_{Te}^2}B\right]. \quad (B.1)$$

Because of $\mathbf{u}=(0,0,u_z)$, we obtain that

$$F_{ei} = -\frac{3N_e N_i Z_i^2 e^4 u_z \ln\Lambda}{16\pi^2\varepsilon_0^2 k_B T_e}\frac{\rho_{\alpha,r,\sigma}}{X_{r,\sigma}^{3/2}v_{Te}^3}\int_{-v_{\max}}^{v_{\max}}\frac{v_z^2}{v^3}\left[AB^{-\sigma-1}\left(\frac{\mathbf{v}^2}{X_{r,\sigma}v_{Te}^2}\right)^r\frac{2\sigma(1+r)}{X_{r,\sigma}(\sigma-1)}-\frac{4\alpha\mathbf{v}^2}{v_{Te}^2}B\right]d\mathbf{v}$$

$$= -\frac{N_e N_i Z_i^2 e^4 u_z \ln\Lambda}{4\pi\varepsilon_0^2 k_B T_e}\frac{\rho_{\alpha,r,\sigma}}{X_{r,\sigma}^{3/2}v_{Te}^3}\int_0^{v_{\max}} v\left[AB^{-\sigma-1}\left(\frac{v^2}{X_{r,\sigma}v_{Te}^2}\right)^r\frac{2\sigma(1+r)}{(\sigma-1)X_{r,\sigma}}-\frac{4\alpha v^2}{v_{Te}^2}B\right]dv. \quad (B.2)$$

When we let $v_{\max}=\infty$, we have that

$$F_{ei} = -\frac{N_e N_i Z_i^2 e^4 u_z m_e^{1/2} \ln\Lambda}{4\pi\varepsilon_0^2(k_B T_e)^{3/2}}\frac{\rho_{\alpha,r,\sigma}}{X_{r,\sigma}^{3/2}}, \quad (B.3)$$